PHYSICAL SCIENCES    Physics

# Pressure induced superconductivity in $CaC_2$

Yan-Ling Li,[a,b] Wei Luo,[b,c] Zhi Zeng,[d] Hai-Qing Lin,[e] Ho-kwang Mao,[f] and Rajeev Ahuja,[b,c]

[a] *School of Physics and Electronic Engineering, Jiangsu Normal University, 221116, Xuzhou, P. R. China*

[b] *Condensed Matter Theory Group, Department of Physics and Astronomy, Uppsala University, P.O. Box 516, SE-751 20 Uppsala, Sweden*

[c] *Applied Material Physics, Department of Materials Science and Engineering, Royal Institute of Technology (KTH), SE-100 44, Stockholm, Sweden*

[d] *Key Laboratory of Materials Physics, Institute of Solid State Physics, Chinese Academy of Sciences, Hefei 230031, P. R. China*

[e] *Beijing Computational Science Research Center, Beijing 100084, P. R. China*

[f] *Geophysical Laboratory, Carnegie Institution of Washington, Washington, DC 20015, USA*

Carbon can exist as isolated dumbbell, one-dimensional (1D) chain, 2D plane, and 3D network in carbon solids or carbon-based compounds, which attributes to its rich chemical binding way, including $sp$-, $sp^2$, and $sp^3$-hybridized bonds. $Sp^2$ hybridizing carbon always captures special attention due to its unique physical and chemical property. Here, using evolutionary algorithm in conjunction with *ab initio* method, we found that under compression, dumbbell carbon in $CaC_2$ can be polymerized firstly into one-dimensional chain and then into ribbon and further into two-dimensional graphite sheet at higher pressure. The *C2/m* structure transforms into an orthorhombic *Cmcm* phase at 0.5 GPa, followed by another orthorhombic *Immm* phase, which is stabilized at wide pressure range of 15.2-105.8 GPa and then forced into $MgB_2$-type phase with wide range stability up to at least 1 TPa. Strong electron-phonon coupling λ in cold compressed $CaC_2$ is found, in particular for *Immm* phase, which has the highest λ value (0.562-0.564) among them, leading to its high superconducting critical temperature $T_c$ (7.9~9.8 K), which is comparable with 11.5 K value of $CaC_6$. Our research results show that calcium can not only stabilize carbon $sp^2$ hybridization at larger range of pressure but make them present superconducting behavior, which would further ignite experimental and theoretical research interests on Alkaline-Earth metal carbides to uncover their peculiar physical properties under extreme conditions.

High pressure, as a crucial thermodynamic parameter, has been emerging as a powerful tool to investigate physical and chemical behavior of materials especially to explore the evolution of planets, and synthesize or design materials with peculiar properties such as superhardness and superconductivity (1-5). Recently, Ca and C, as ordinary elemental solids, have been studied extensively in physical, chemical and material science fields due to their interesting structural property when applying pressure (6-10). These peculiar physical properties of compressed Ca and C solids have motivated our attention on Ca-based dicarbide $CaC_2$. $CaC_2$ (11,12) at ambient pressure, experimentally, four temperature-induced modifications were reported, that are the well-known tetragonal room temperature modification $CaC_2$ I (space group *I4/mmm*, Z=2), two low-temperature monoclinic modifications $CaC_2$ II (*C2/c*, Z=4) and $CaC_2$ III (*C2/m*, Z=4), and cubic high-temperature modification $CaC_2$ IV (*Fm-3m*, Z=4) (12). Theoretically, orthorhombic $CaC_2$ *Immm* (Z=2, referred as *Immm*-1 below) structure, monoclinic $CaC_2$ *C2/m* (Z=2, referred as *C2/m*-1 below) structure, and trigonal $CaC_2$ *R-3m* (Z=1) structure, were predicted by Kulkarni *et al* using a global exploration method based on simulated annealing scheme (13). In all structures reported above, carbon atoms form isolated dumbbells. Since both Ca and C solids tend to form C-atom network structures at ambient and high pressure (6-10), it is reasonable and interesting to investigate pressure-induced possibility of polymerization of isolated $C_2$ dumbbells in $CaC_2$. To our knowledge, there are few studies on compressed $CaC_2$ so far. Kulkarni *et al* observed that *C2/m*-1 structure transforms into *R-3m* structure at about 24 GPa using LDA or 34 GPa using B3LYP (13). Srepusharawoot *et al* studied pressure-induced carbon network formation for five Alkaline-Earth metal dicarbides (AEMDs) using *ab initio* random searching method (14). Very recently, structural and vibrational properties of AEMD $BaC_2$ at ambient temperature and high pressure up to 40 GPa were reported experimentally (15). Amorphous states were observed which can be viewed as a precursor for the formation of carbon networks embedded in a metal framework.

Covalent bonding and ionic bonding co-exist in AEMDs. There is a strong covalent bonding in $C_2$ units, but ionic in between Ca and C. Up to now, there are many open questions for compressed AEMD: 1) how $C_2$ dumbbell evolved after applying pressure? 2) Could it be metalized under reasonable pressure? 3) Is metallic high pressure phase superconductor? Here, we select $CaC_2$ as an example to address these problems.

The structural properties of $CaC_2$ at high pressure were investigated by using global structural searching scheme based on evolutionary algorithm (EA) (16, 17) in combination of first-principles total energy calculations. At high pressure, it was found that $CaC_2$ adopt the layered structures, in which carbon atoms firstly form chains and then ribbons or graphene sheets separated by Ca atomic layers. Three of the predicted high pressure phases, two of them have orthorhombic symmetry (*Cmcm* and *Immm*) and third one is well-known $MgB_2$ type structure with *P6/mmm* symmetry. Calculated vibrational properties are similar as those observed in $CaC_6$ (18). The electronic structure calculations show that all the three predicted high pressure phases are metallic. Finally, phonon-mediated superconducting behavior of three new high pressure phases of $CaC_2$ was revealed by exploring electron-phonon coupling.

Knowledge the structural information is the first step toward understanding the physical properties of any material. To get the most stable structure of $CaC_2$, we searched for its lower enthalpy's phase based on EA in combined with VASP package. We performed variable-cell structure prediction simulations using USPEX code for $CaC_2$ containing two, three, and four molecules in the simulation cell at 5, 20, 40, 80, 120, 200 GPa, 300 GPa, 500GPa, 700GPa, and 1TPa, respectively. During the evaluation, the total energy was calculated via VASP package. For comparison, the structures discussed in $BeC_2$, $MgC_2$, and $BaC_2$ experimentally and theoretically, are also considered in our calculations. The enthalpies per chemical formula unit vs. pressure curves of selected structures are plotted in Fig. 1. Considering that *I4/mmm*, *C2/m*, *C2/c*, *C2/m*-1, and *Immm*-1 structures hold very close enthalpy, their enthalpies vs. pressure curves were specially given in Fig. 1 b), in which *I4/mmm*, *C2/c*, and *C2/m* refer to three modifications determined at different temperature by experiments, while *C2/m*-1 and *Immm*-1 represent lower enthalpy structures suggested by a global exploration method based on simulated annealing (13). From Fig.1, one can see that, for compressed $CaC_2$, the monoclinic *C2/m* structure is the most stable structure below 0.5 GPa. An orthorhombic *Cmcm* phase takes over the pressure range from 0.5 to 15.2 GPa. Another orthorhombic *Immm* structure is the most stable structure over the wide pressure range of 15.2 to 105.8 GPa, followed by a $MgB_2$ type structure (space group *P6/mmm*) at 105.8 GPa above. Among these, *Cmcm*, *Immm* and *P6/mmm* phases are metallic. Our total energy calculations rule out other competitive low-enthalpy structures which exist in MgC$_2$, CaC$_2$, and BaC$_2$, including *I4$_1$/amd* (16), *P4$_2$/mnm* (20), *P4$_3$32* (19), *I4/mmm*, *C2/m*-1, *Immm*-1, *P6$_3$/mmc*, and *R-3m* (13). Our calculations show that the previous *R-3m* structure of CaC$_2$ suggested theoretically (13) is not favorable one (see Fig. 1). In addition, *Cmmm* structure, a meta-stable phase, has comparative enthalpy with *Immm* and *P6/mmm* phases at large pressure range due to their similar structural properties (atomic layered arrangement). The optimized structural parameters at different pressure are listed in Table I. For *Cmcm* phase, the equilibrium lattice constants at 4 GPa are *a*=3.6822 Å, *b*= 8.6324 Å, and *c*=4.7360 Å. In this structure, four calcium atoms lie in the Wyckoff 4*c* site and eight carbon atoms occupy 8*f* site, in which calcium atoms construct cylinder with hexagon cross section, centered by one dimensional armchair carbon chain along z direction (see Fig.2 c). The equilibrium lattice constants of *Immm* phase at 15.2 GPa are *a*=7.0623 Å, *b*=2.6317 Å, and *c*=6.2697 Å. Four calcium atoms hold the Wyckoff 4*e* sites and eight C atoms occupy two in-equivalent Wyckoff 4*i* (referred as C1) and 4*j* (referred as C2) sites. It is worth to note that *Immm* structure suggested here differs clearly from *Immm*-1 structure predicted by simulated annealing (13). *Immm*-1 phase includes two molecules per unit cell, in which carbon atoms form isolated dumbbells whereas our stable *Immm* structure contains four molecules per unit cell, in which carbon atoms were polymerized into nanoribbon with six-membered ring. From Fig.2 d, one can see that carbon nanoribbon, forming quasi-one dimensional (1D) carbon ribbons, lies in the center of cylinder constructed by calcium atoms. For *P6/mmm* phase, a MgB$_2$-type structure, including only one molecule per unit cell, the equilibrium lattice parameters at 105.8 GPa are *a*=2.5412 Å, and *c*=3.6864 Å. One calcium atom exists in 1*a* site and two carbon atoms lie in 2*d* sites. Structurally, *P6/mmm* phase of CaC$_2$ consists of hexagonal honeycombed layers (referred as graphene sheet) of carbon atoms separated by planes of Ca atoms, with the calcium atoms centered above and below the carbon hexagons (See Fig. 2). Also we have noticed that pressure-induced phase transition in CaC$_2$ is first-order phase transition because of obvious volume's change at phase transition point. Under chemical pre-compression and external compression, dumbbell-type carbon is polymerized firstly into ordered armchair chain and further into quasi-1D well-ordered nanoribbon and eventually into two dimensional graphite sheets (graphene). Besides this, we did explore possibility of three-dimensional network carbon under higher pressure until 1 TPa based on EA. Surprisingly, all performed calculations always point to MgB$_2$-typre (*P6/mmm*) structure. Phonon calculation further confirms its dynamical stability.

It will be constructive to look at the change of carbon-carbon bonding with increasing pressure to study the structural phase transformation mechanism. From low pressure phase to high pressure phase, C-C bonding behavior reveals obvious change due to external pressure and internal chemical pre-compression. Applying pressure to CaC$_2$, *I4/mmm* structure with isolated dumbbell space orientation along *z*-axis, which is stable at room temperature observed experimentally, is not energetically favorable compared to those (such as, *C2/m*, *C2/c*, *C2m*-1 and *Immm*-1, see Fig. 1) which possess isolated dumbbell space orientation keeping an angle with z-axis. With increasing pressure, the distance between isolated dumbbell decreases, resulting in carbon atomic chain formed at lower pressure (see *Cmcm* structure). Upon further pressure, carbon atomic chain tends to well-organized arrangement, making orthorhombic *Cmcm* phase being transformed into an orthorhombic *Immm* phase. In *Immm* structure, there are two different bonding lengths (referred as $d_1$ and $d_2$) in ribbon with six-membered carbon ring (see Fig. 3). At the pressure of 10 GPa, two among six C-C bondings ($d_1$) hold length of 1.4747 Å, and four of them ($d_2$) have length of 1.5246 Å. Under further compression, the difference ($\Delta d$) between $d_1$ and $d_2$ begin to decrease smoothly (see Fig. 3 a). At the same time, the distance ($d_3$) between neighbor ribbons begins to decrease (see Fig. 3). In other words, for *Immm* phase, the difference of carbon-carbon bond length gradually decreases with increasing pressure until it disappears at 150 GPa which results in a formation of a regular hexagonal ring. This kind of pressure-induced structural modification indicates the feasibility of graphite sheet (i.e. graphene sheet) formed in compressed $CaC_2$, as is observed in *P6/mmm* structure.

Due to the occurrence of graphite sheet between neighbor Ca atomic layers, *P6/mmm* phase could be one of graphite intercalation compounds (GICs). The properties of GIC family have been studied extensively using a variety of different experimental techniques (21-25). Especially, the discovery of superconductivity in $CaC_6$ at the enhanced transition temperature (11.5 K) (26, 27) excited intense research interests on GICs experimentally (28-29). The C-C bond length in *P6/mmm* structure is 1.4672 Å at 105.8 GPa, which is slightly longer than that (1.42 Å) in graphite. Also, we can see that *Immm* phase has similar structural property as $CaC_6$ (all the calcium atoms lie above and below of the center of carbon six-membered ring). This peculiar structural behavior indicates that compressed $CaC_2$ would present similar physical properties as $CaC_6$, such as superconductivity.

It is essential to check mechanical and dynamical stability of low enthalpy phases by means of elastic constants and phonon spectrum. There are thirteen, nine, and five independent elastic constants for monoclinic, orthorhombic, and hexagonal structures, respectively. The calculated elastic constants for four low enthalpy phases of $CaC_2$ were presented in Table 1. For hexagonal symmetry, $c_{66}$ equals to $(c_{11}-c_{12})/2$. From Table 1, it is obvious that elastic constants satisfy mechanical stable criterion, implying their stability mechanically. Additionally, bulk modulus *B*, shear modulus *G*, and Young's modulus *E* were obtained using Voigt's formulas. For phase *C2/m*, the calculated *B*, *G* and *E* are 55, 24, and 63 GPa. Small *B* value indicates its strong compressibility. We have noticed that volume collapse 19.3 % at phase transition point from *C2/m* to *Cmcm*. Strong compressibility attributes to highly ionic bonding in between calcium and carbon. For hexagonal *P6/mmm* phase, high elastic modulli originate from strong covalent bonding between carbon atoms.

Phonon calculations show that *Cmcm*, *Immm* and *P6/mmm* phases are stable dynamically at their stable pressure ranges. Phonon dispersions and partial phonon density of states (PPHDOS) of phases *Cmcm*, *Immm,* and *P6/mmm* are shown in Fig. 4. The maximum optical branch frequencies are 1390.4 $cm^{-1}$ for *Cmcm* phase at 1

GPa, 1129.7 cm$^{-1}$ for *Immm* phase at 15.2 GPa and 1443.9 cm$^{-1}$ for *P6/mmm* at 105.8 GPa, which are lower than the value 1860 cm$^{-1}$ (30) of room temperature phase *I4/mmm* due of obvious structural difference. From PPHDOS, one can conclude that calcium atom dominates low frequency region, while carbon atoms contributes to high frequency region. In Fig. 4, flat bands along peculiar direction in BZ indicate layered structural property for three new phases. For *Cmcm* structure, the phonon dispersions possess two frequency gaps (48.4 and 171.6 cm$^{-1}$), which originates from the stiffening of the carbon chains.

*P6/mmm* phase contains interesting graphene sheets, so we discuss only phonon behavior of *P6/mmm* phase in the following. For *P6/mmm* phase, PPHDOS calculation shows that the Ca-related phonons occur at the lower frequencies typically below 491 cm$^{-1}$ since the Ca atoms are much heavier and more weakly bonded than the C atoms (see Fig. 4). In detail, for *P6/mmm* phase, Ca$_y$ modes mainly occupy in below 400.3 cm$^{-1}$, while Ca$_x$ and Ca$_z$ modes dominate the spectrum range from 400.3-700.6 cm$^{-1}$. The graphene sheets in *P6/mmm* phase have strong planer bonding, indicating that the out-of-plane C$_z$ modes fall between 300.2-667.2 cm$^{-1}$ while the in-plane C$_{xy}$ modes are mainly in higher frequency and dominate the frequencies above 700.6 cm$^{-1}$ (see Fig. 4). So there is strong coupling between Ca$_{xz}$ phonons and C$_z$ phonons. The observed carbon atomic vibration behavior here has been determined in well-known Ca-based carbide CaC$_6$ experimentally. The most important issue is that the calculated phonon frequency windows of C$_z$ modes and C$_{xy}$ modes in CaC$_2$ are very similar as that observed in CaC$_6$, implying that *P6/mmm* phase of CaC$_2$ has similar phonon-related property as CaC$_6$. The partial atomic phonon distribution of *Immm* phase is similar as *P6/mmm* phase. All three phases possess soft phonon modes at peculiar high symmetry point direction in BZ. In particular, *Immm* phase has more soft modes at low frequency region along G-R, G-X, and G-T directions in BZ (Fig.4 b), which serves as evidence of strong electron-phonon coupling in *Immm* phase (discuss later).

Electronic structure calculations show that *C2/m* phase is semiconductor (see Fig. S1), while *Cmcm*, *Immm* and *P6/mmm* structures are metallic. From projected density of states (PDOS) given in Fig. 5, one can see that three metallic phases have consistent electronic distributions. C-*p* electrons dominate in the wide energy range of covalence bands and strongly hybridize with Ca-*d* electrons near the Fermi level (i.e., Ca-*d* electrons and C-*p* electrons dominate Fermi level). On the other hand, there are few C-*s* electrons and Ca-*s* electrons near Fermi level, denoting a pressure driven charge transfer from *s* electrons to *p* or *d* electrons, which is confirmed by atomic Mulliken population analysis shown in Table S2. According to this, one can conclude that the majority of Ca-4*s* electrons transform to Ca-*d* orbital, while the remaining into C atom. For C atom, charge transfer is clearly from 2*s* orbital to 2*p* orbitals. The pressure-driven *s-d* charge transfer makes newly predicted phases being favorable ones energetically. In addition, we considered pressure effect by taking *Immm* phase as an example. From Fig. S2, it is obvious that pressure broaden the bands, making conduction band shift to higher level and covalence band shift to lower level. Additionally, projected density of states calculations show that near the Fermi level, the hybridization between C-$p$ states and Ca-$d$ states is enhanced with increasing pressure.

Observed soft phonon modes, similar vibrational properties as $CaC_6$, and flat energy bands near Fermi level in *Immm* phase arouse our interest to explore its superconducting behavior. The calculated spectral function $\alpha^2F(\omega)$ and integrated $\lambda(\omega)$ of *Immm* and *P6/mmm* phases of $CaC_2$ at different pressure were plotted in Fig. 6 (For the case of *Cmcm* phase, see Fig. S3). For *Immm* phase, at 17 GPa, the vibrations below 587.2 cm$^{-1}$ provide the major contribution to $\lambda$, accounting for 66.7 % of total $\lambda$ value (about 0.565). The phonons below 150.1 cm$^{-1}$, which mostly attributes to Ca atom, contribute only 2.9% of total $\lambda$ value. The phonons between 150.1 and 280.2 cm$^{-1}$, origins mostly from Ca-yz modes (along Ca-C1 bonding direction) as well as a small quantity of C1-yz modes (along Ca-C1 bonding direction), contribute 39.6 % of the total $\lambda$. The phonons between 280.2 and 383.6 cm$^{-1}$, mostly origins from Ca-x modes (parallel to carbon ribbon plane) as well as a small quantity of C1-yz modes (along Ca-C1 bonding direction), contribute 10.1 % of the total $\lambda$. The phonons between 383.6 and 587.2 cm$^{-1}$, mostly origins from C2-yz modes (along Ca-C2 bonding direction), contribute 8.7 % of the total $\lambda$. To summarize, strong electron-phonon coupling which is necessary for superconductivity in *I/mmm* phase of compressed $CaC_2$ is due to the phonons from Ca and C1 atoms together with electrons from the Ca-$d$ and C-$p$ states. From Fig. 6, one can see that there is obvious difference between *Immm* and *P6/mmm* in mechanism of electron-phonon coupling. For *P6/mmm* phase, at 106 GPa, the vibrations below 700.6 cm$^{-1}$ provide the major contribution to $\lambda$ (about 0.47). The low-frequency phonons (below 300.2 cm$^{-1}$), which mostly involve the Ca atoms yield only 6.6 % of total $\lambda$ value. The phonons between 300.2 and 700.6 cm$^{-1}$ contribute 75.4 % of the total λ. The frequencies at 700.6 cm$^{-1}$ above, that is, the in-plane vibration modes from carbon atoms, only contribute 18 % of the total $\lambda$. These results combined with PPHDOS indicate that out-of-plane C$_z$ modes in the *P6/mmm* structure dominate superconductivity in phase *P6/mmm* of CaC$_2$, due to the prominent contributions to the electron-phonon interaction. Phonons from out-of-plane C$_z$ modes together with the electrons from the Ca-*d* and C-*p* states provide the strong electron-phonon coupling necessary for superconductivity in phase *P6/mmm* of compressed CaC$_2$.

The Allen and Dynes modified formula (31) was used to estimate the superconducting transition temperature $T_c$ from the value of $\lambda$ determined above. Taking a typical value of 0.115 for the effective Coulomb repulsion parameter $\mu^*$, (which is thought to be able to get $T_c$ in agreement with 11.5 K (26, 27) of experiment in CaC$_6$ (32)), we calculated $T_c$ of three phases for CaC$_2$ and $T_c$'s dependence on pressure. Calculated $T_c$s and logarithmic phonon momentum ω$_{log}$ vs pressure curves were presented in Fig. 7. Logarithmic phonon momentum ω$_{log}$ increases with pressure increase. Among three phases, *Immm* has the strongest electron-phonon coupling (0.564, see inset figure in Fig. 7) and so have the highest $T_c$. At wide range of pressure, *Immm* phase of CaC$_2$ has comparative superconducting critical temperatures (from 7.9 K at 43 GPa to 9.8 K at 95 GPa) with 11.5 K value (26, 27) of $CaC_6$.

The structural, dynamical, and electronic properties of compressed $CaC_2$ were systemically investigated up to 1TPa. Three stable high pressure phases, *Cmcm*, *Immm* and $MgB_2$-type structures, were predicted by using *ab initio* EA. The carbon atomic arrangement form from 'isolated' dumbbell to 1D chain to quasi-1D ribbon to 2D plane is observed. Phonon calculations have shown their dynamical stability at the dominating pressure range. Strong electron-phonon coupling between Ca-*d* and C-*p* electrons and Ca-yz and C1-yz phonons (vibration along Ca-C1 bonding direction) dominates superconductivity of phase *Immm*, while strong electron-phonon coupling between Ca-*d* electrons and carbon out-of-plane phonons is responsible for superconductivity of phase *P6/mmm*. The predicted evolution of carbon from 'local' dumbbell to 2D graphene sheet and high superconducting critical temperature in compressed $CaC_2$ would stimulate further experimental and theoretical studies on Alkaline-Earth metal carbides.

## Methods

*Ab initio* EA, designed to search for the structure possessing the lowest free energy at given pressure and temperature conditions, has been employed using USPEX code (16,17). The structural and electronic properties of $CaC_2$ over a wide range of the pressure were performed using DFT as implemented in Vienna *ab initio* simulation package (VASP)(33), employing the projected augmented wave (PAW) pseudopotential included in the released pseudopotential library (34,35) where $2s^22p^2$ and $3s^23p^64s^2$ are treated as valence electrons for C and Ca atoms, respectively. Under the PAW approach, Perdew, Burke, and Ernzerhof's exchange correlation functional (36) was chosen for both Ca and C. For C atom, we select hard pseudopotential to carry out our calculations since hard pseudopotential was thought to be more suitable for high pressure research. When searching the stable structures, we performed calculations with relaxation of cell volume, cell shape, and ionic positions. Forces on the ions were calculated through derivatives of the free energy with respect to the atomic positions, including the Harris-Foulkes like correction. All possible structures were optimized using conjugate gradient scheme. For the searches, we used a plane-wave basis set cutoff of 700 eV and performed the Brillouin zone integrations using a coarse k-point grid. The most interesting structures were further relaxed at a higher level of accuracy with a basis set cutoff of 1000 eV and a k-point grid of spacing $2\pi \times 0.018$ Å$^{-1}$. Iterative relaxation of atomic positions was stopped when all forces were smaller than 0.001 eV/ Å.

The dynamical and superconducting properties were calculated in terms of the Quantum-Espresso package (37) using Vanderbilt-type ultra-soft pseudopotentials (38) with cutoff energies of 50 Ry and 500 Ry for the wave functions and the charge density, respectively. The electronic Brillouin zone (BZ) integration in the phonon calculation was based on a $15\times15\times12$, $15\times15\times15$, and $32\times32\times24$ of Monkhorst-Pack k-point meshes for *Cmc*m, *Immm*, and *P6/mmm* phases, respectively. The electron-phonon coupling was convergent with a finer grid of $64\times64\times64$ k points and a Gaussian smearing of 0.01 Ry. The dynamic matrix was computed based on a $4\times4\times4$ mesh of phonon wave vectors for the *Cmcm* and *P6/mmm* structures and on a $3\times3\times3$ mesh of phonon wave vectors for the *Immm* structure. All calculations were carried out using a primitive cell, which can largely reduce the amounts of computation in comparison with using a unit cell.

This work was supported by the NSFC (11047013), the Priority Academic Program Development of Jiangsu Higher Education Institutions (PAPD), and Jiangsu Overseas Research & Training Program for University Prominent Young & Middle-aged Teachers and Presidents. Part of the calculations was performed at the Center for Computational Science of CASHIPS and the Swedish National Infrastructure for Computing (SNIC). We also thank Swedish Research Council (VR) for financial support.

**Figure Legends**

**Fig. 1.** (Color online) The relative enthalpy per chemical formula molecule as a function of pressure for competing structures. The phase transition series monoclinic *C2/m* phase to orthorhombic *Cmcm* phase (0.5 GPa) to another orthorhombic *Immm* phase (15.2 GPa) to hexagonal *P6/mmm* phase (105.8 GPa) obtained. Further calculations determined the stability of hexagonal phase up to 1 TPa.

**Fig. 2.** (Color online) Phase *C2/m* and three new phases of $CaC_2$. The big and small balls represent calcium and carbon atoms, respectively. One dimensional armchair carbon chains along *c*-axis direction observed in *Cmcm*. One dimensional carbon ribbons and two dimensional graphene sheets observed in *Immm* and *P6/mmm* phases, respectively. In *Immm*, calcium atoms construct cylinder with hexagon cross section, which is arranged periodically, centered by one dimensional carbon ribbon (left of d)).

**Fig. 3.** (Color online) Change of structural property for ribbon formed by carbon atoms in *Immm* phase with increasing pressure. Upon compression, the lengths of two kinds of C-C bonds in ribbon tend to uniform.

**Fig. 4.** (Color online) Phonon spectrum and partial atomic phonon density of states of three new phases for cold compressed $CaC_2$. Soft phonon modes observed in them, in particular, for *Immm* phase.

**Fig. 5.** (Color online) Energy band and Projected Density of States (PDOS) of *Cmcm* at 4GPa, *Immm* at 15.2 GPa and *P6/mmm* at 105.8 GPa. The Fermi level taken as energetic reference point.

**Fig. 6.** (Color online) The Eliashberg phonon spectral function $\alpha^2F(\omega)$ (blank line) and integrated $\lambda(\omega)$ (red line) for *Immm* and *P6/mmm* at four pressure points.

**Fig. 7.** (Color online) Calculated $T_c$s and logarithmic phonon momentum $\omega_{log}$ vs pressure. The inset shows the integrated electron-phonon coupling $\lambda$ as a function of pressure.

Table 1. Independent Elastic constants, bulk, shear, and Young's modulli (unit: all in GPa) of stable phases of $CaC_2$.

| | P | $c_{11}$ | $c_{22}$ | $c_{33}$ | $c_{44}$ | $c_{55}$ | $c_{66}$ | $c_{12}$ | $c_{13}$ | $c_{23}$ | $c_{15}$ | $c_{25}$ | $c_{35}$ | $c_{46}$ | B | G | E |
|---|---|---|---|---|---|---|---|---|---|---|---|---|---|---|---|---|---|
| *C2/m* | 0 | 86 | 98 | 91 | 13 | 47 | 6 | 29 | 58 | 22 | 7 | 9 | -21 | 5 | 55 | 24 | 63 |
| *Cmcm* | 4 | 158 | 188 | 317 | 70 | 72 | 48 | 43 | 68 | 78 | | | | | 115 | 70 | 174 |
| *Immm* | 15.2 | 338 | 650 | 397 | 162 | 79 | 86 | 64 | 59 | 71 | | | | | 197 | 145 | 349 |
| *P6/mmm* | 105.8 | 1056 | | 824 | 373 | | 458 | 141 | 238 | | | | | | 463 | 396 | 924 |

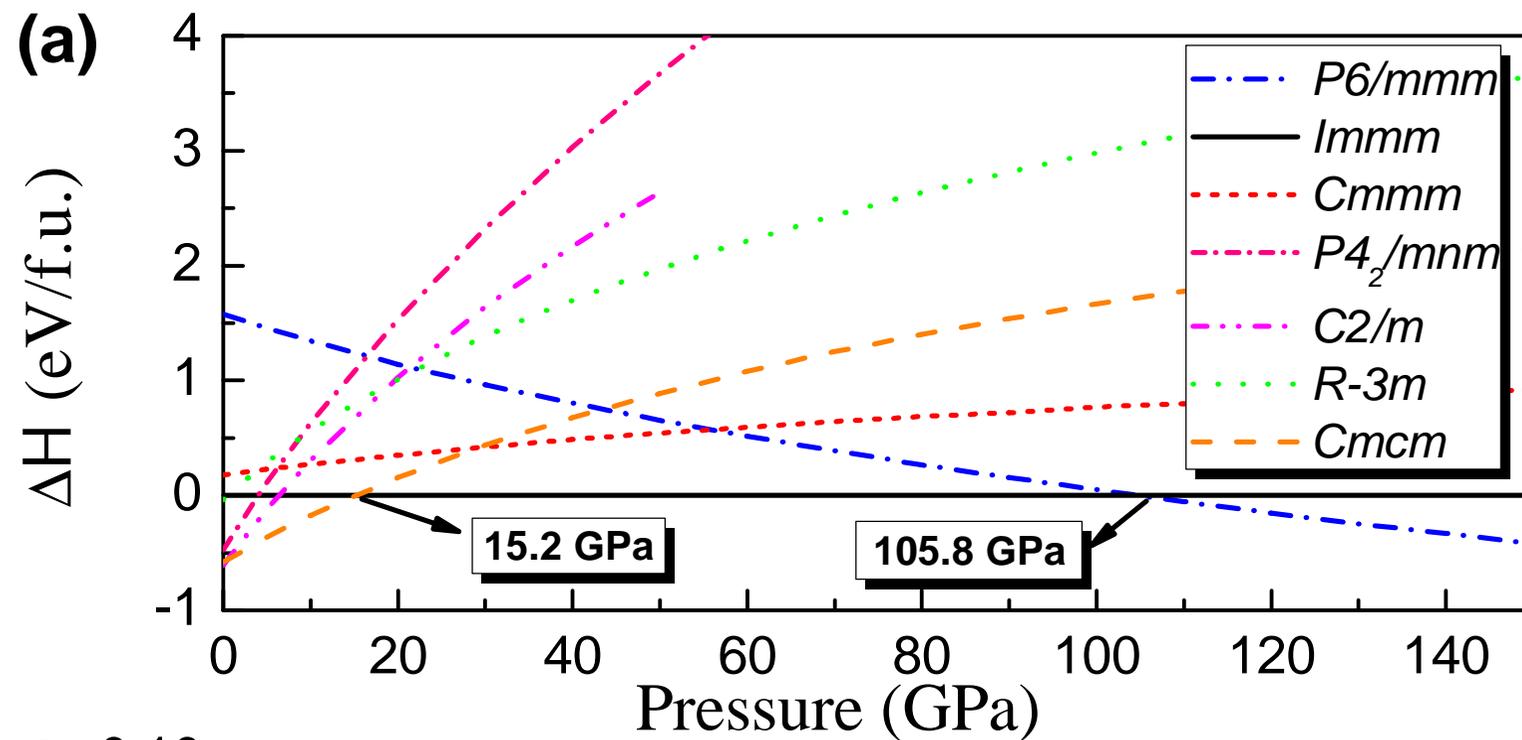
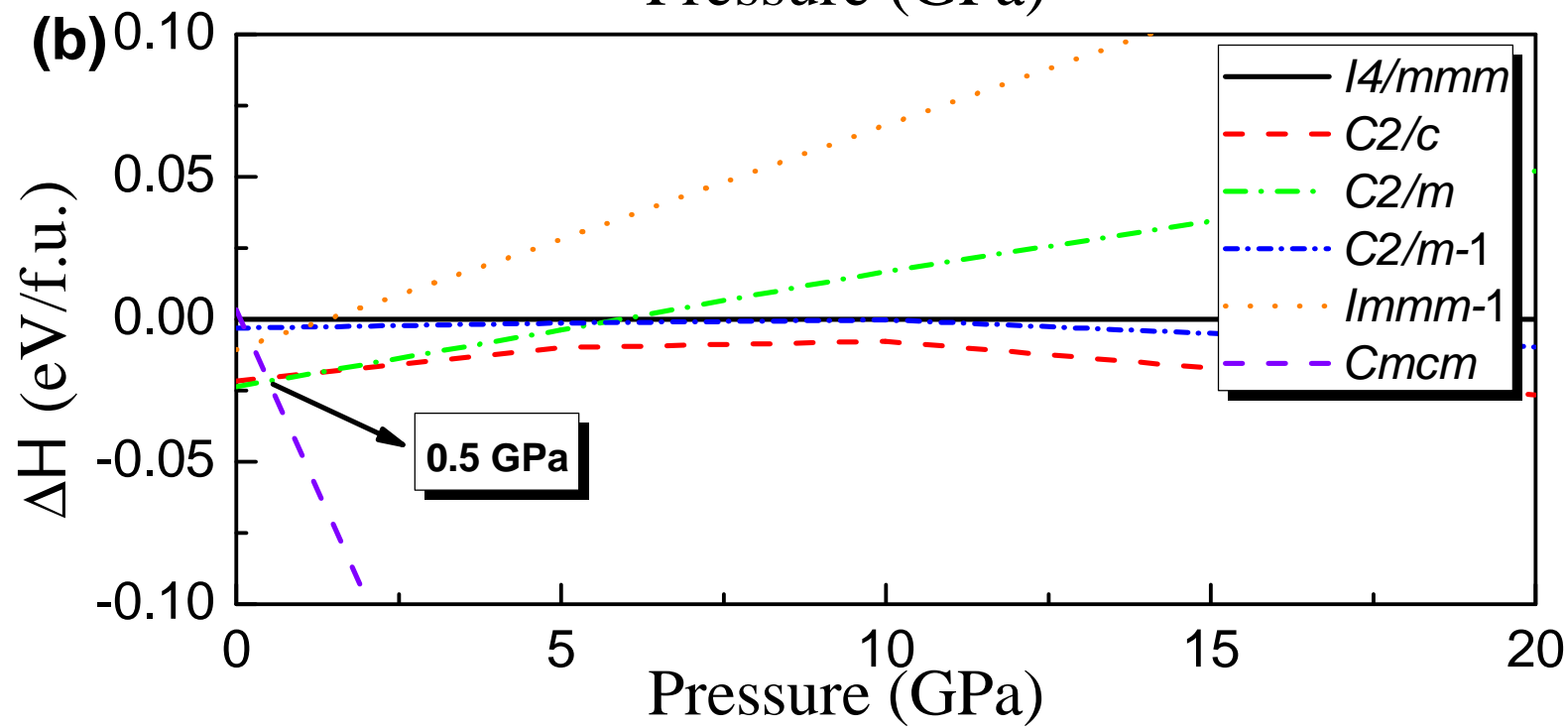

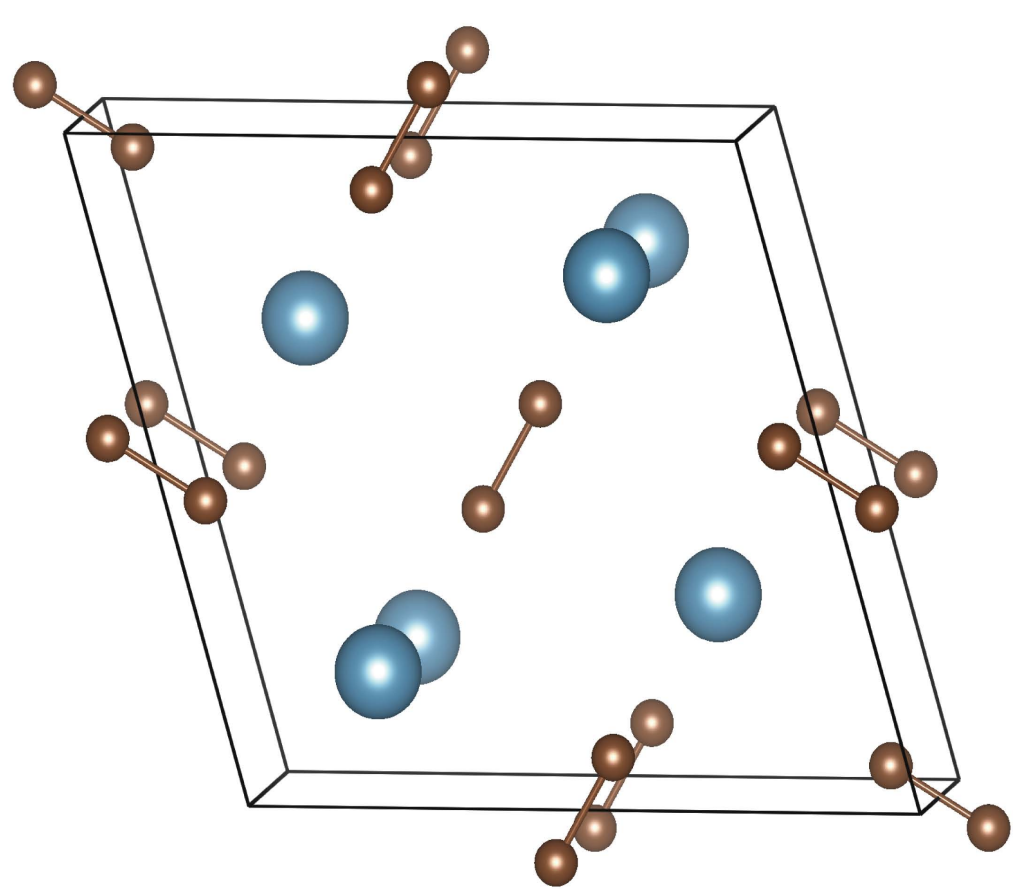
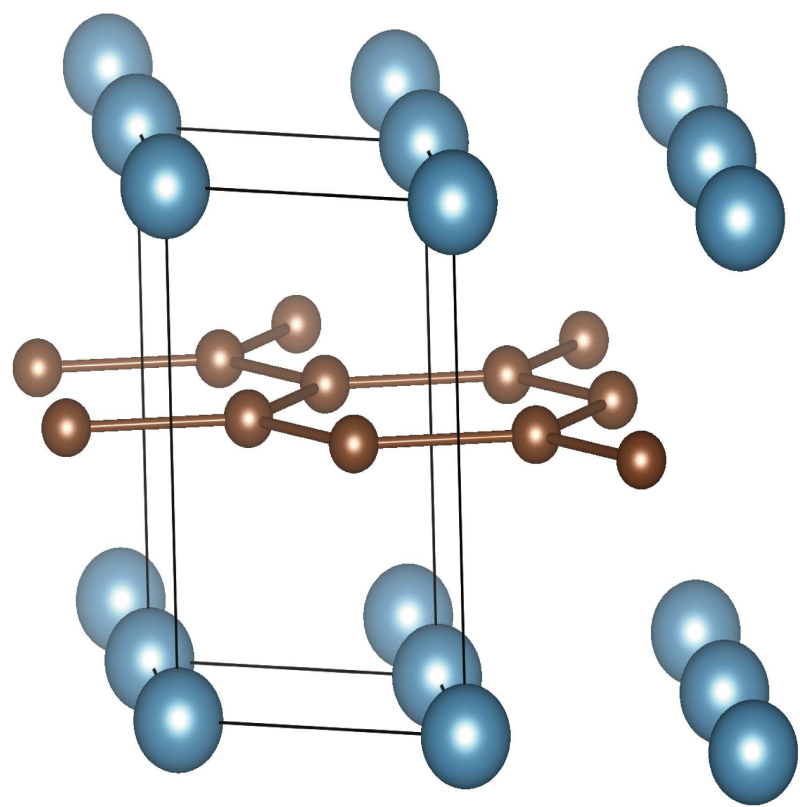

(a) *C2/m*  (b) *P6/mmm*

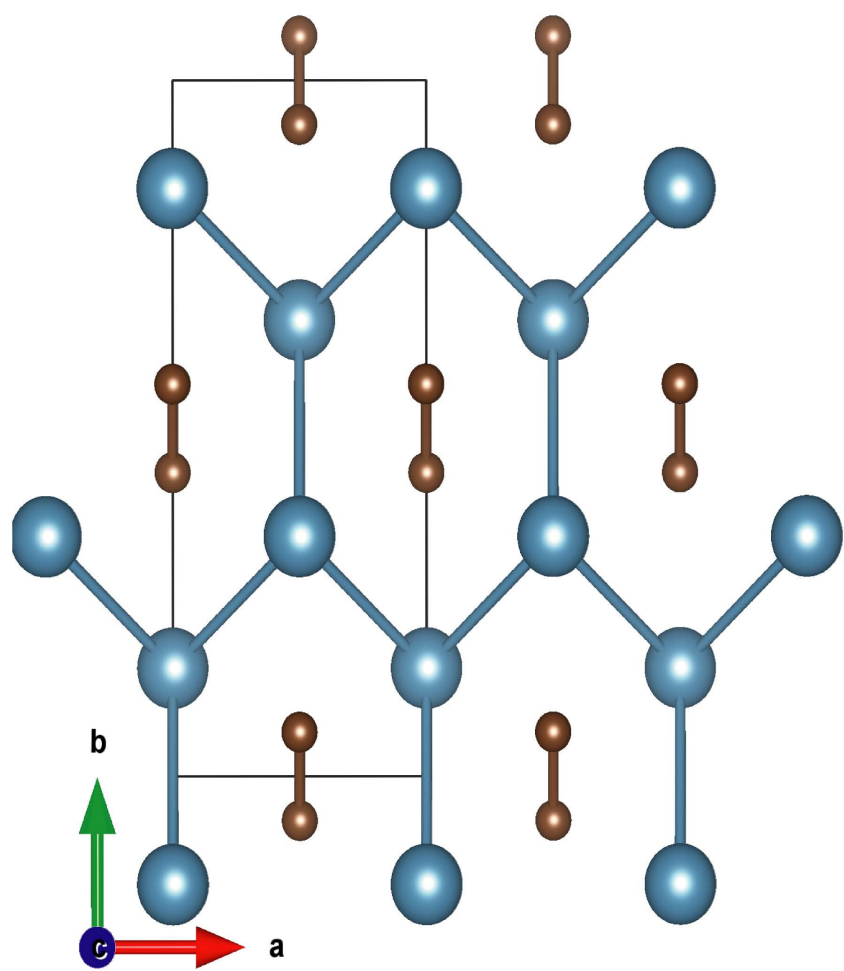
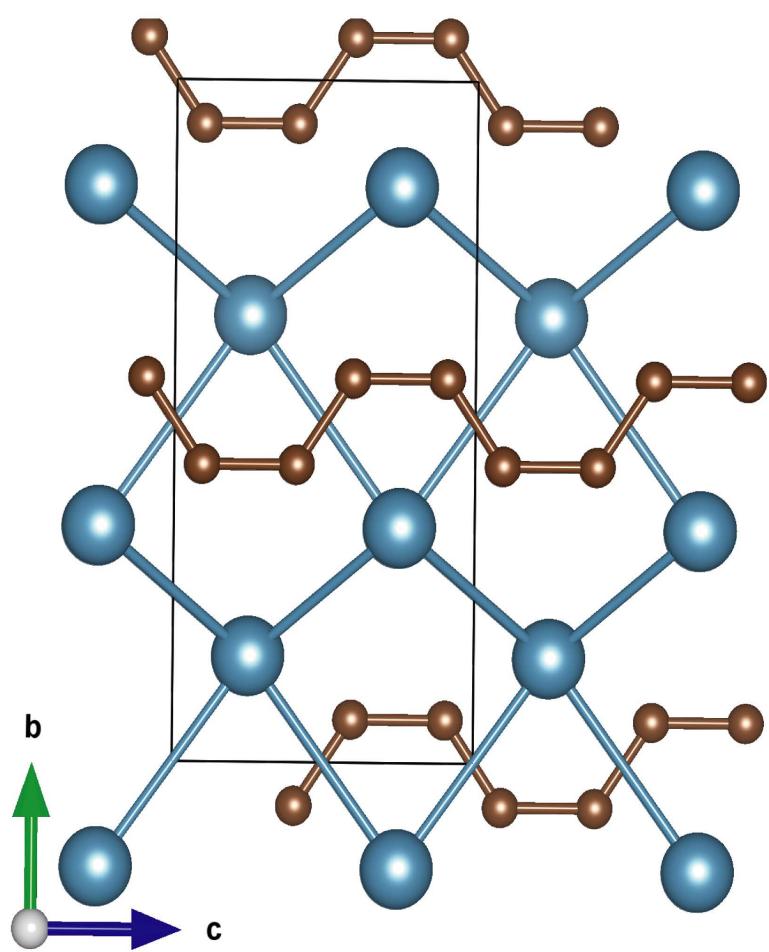

(c) *Cmcm*

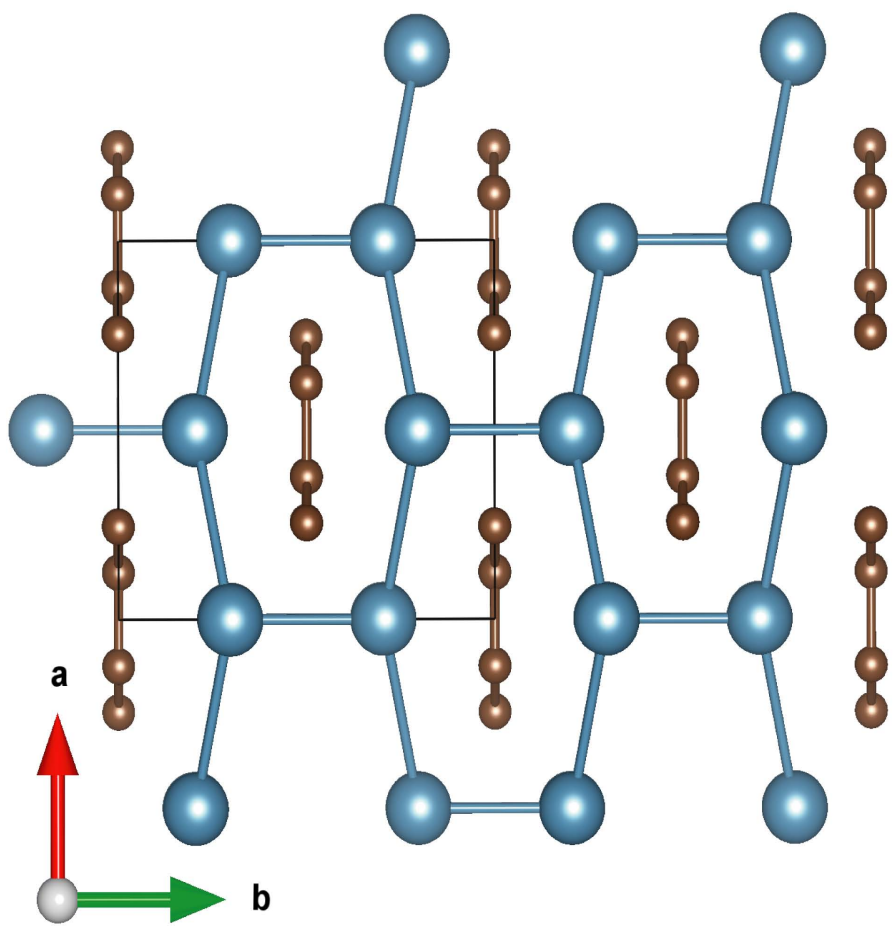
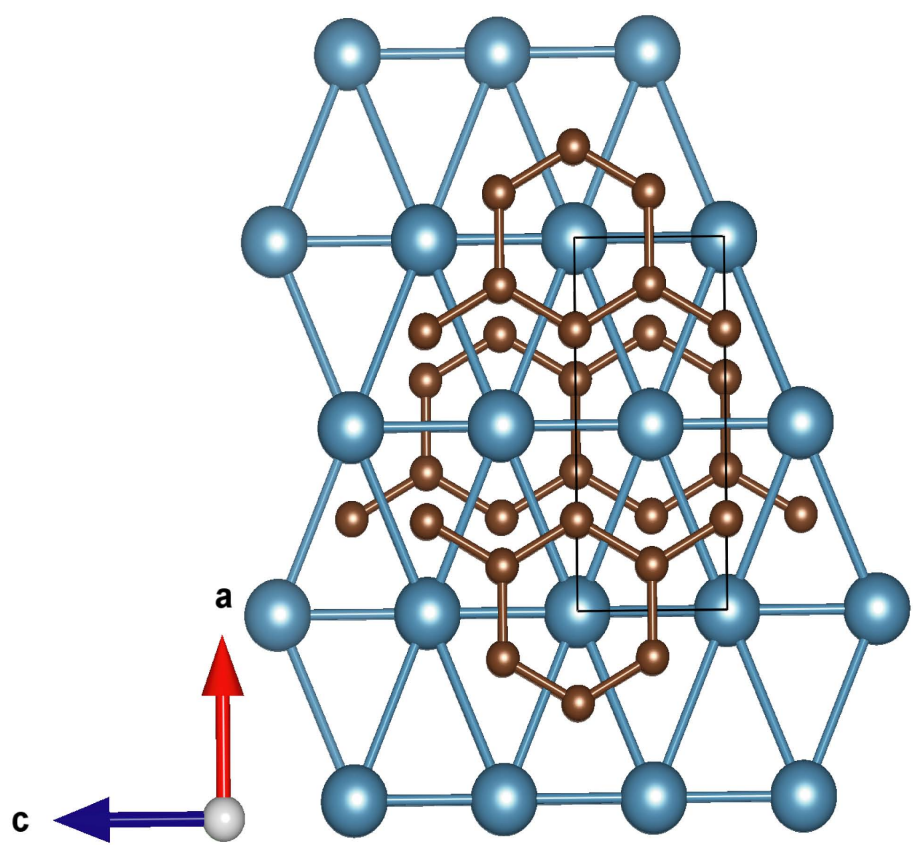

(d) *Immm*

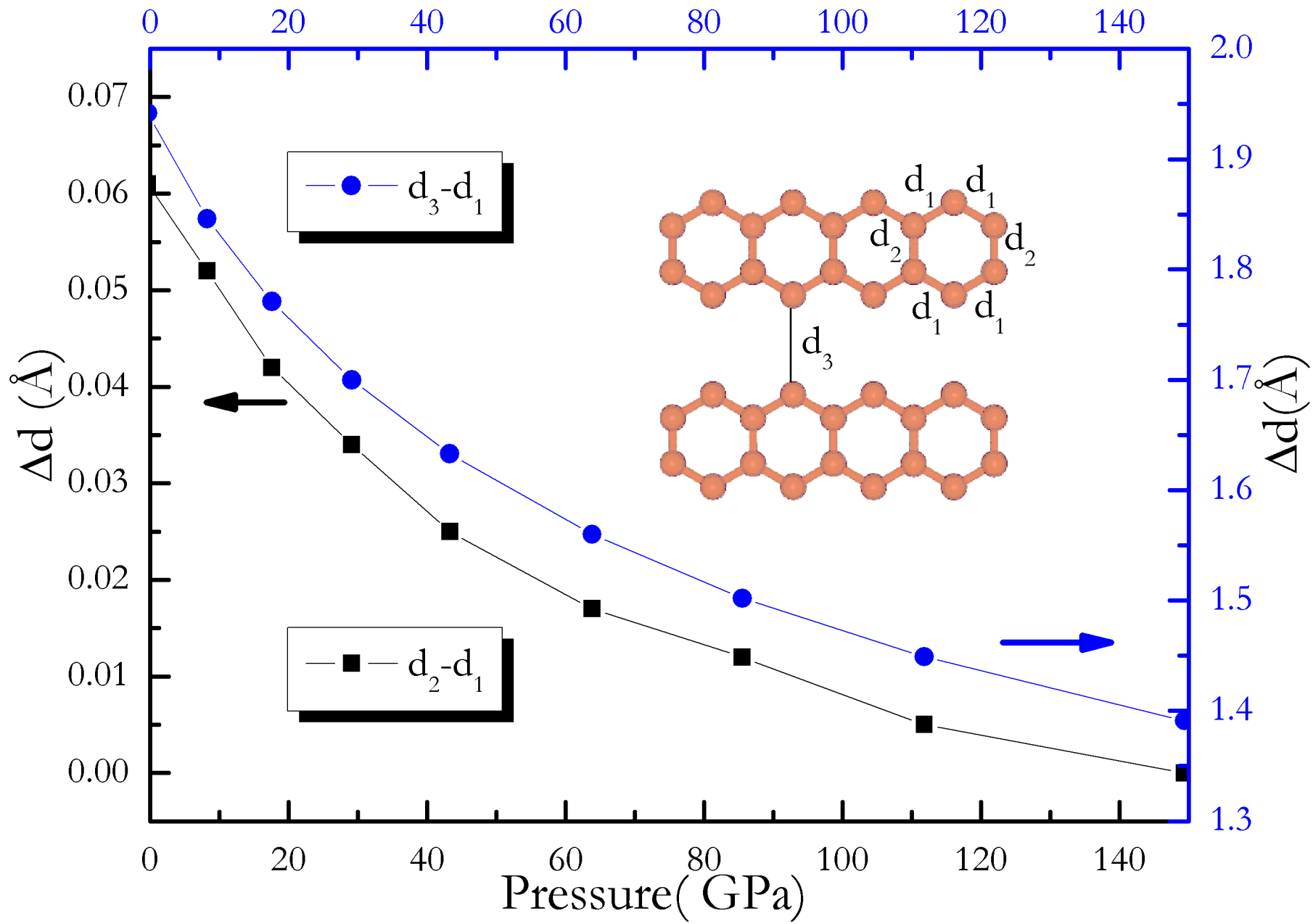

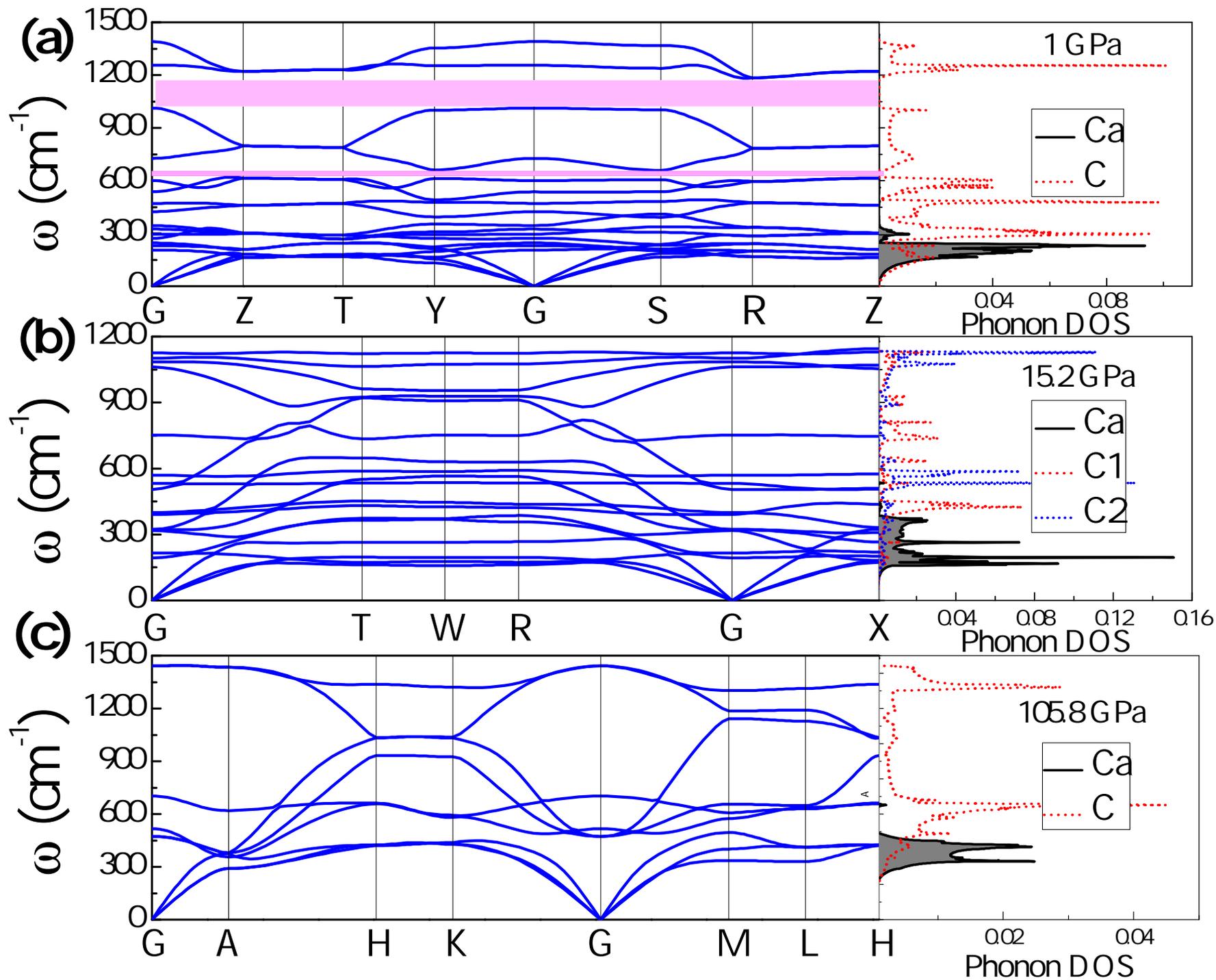

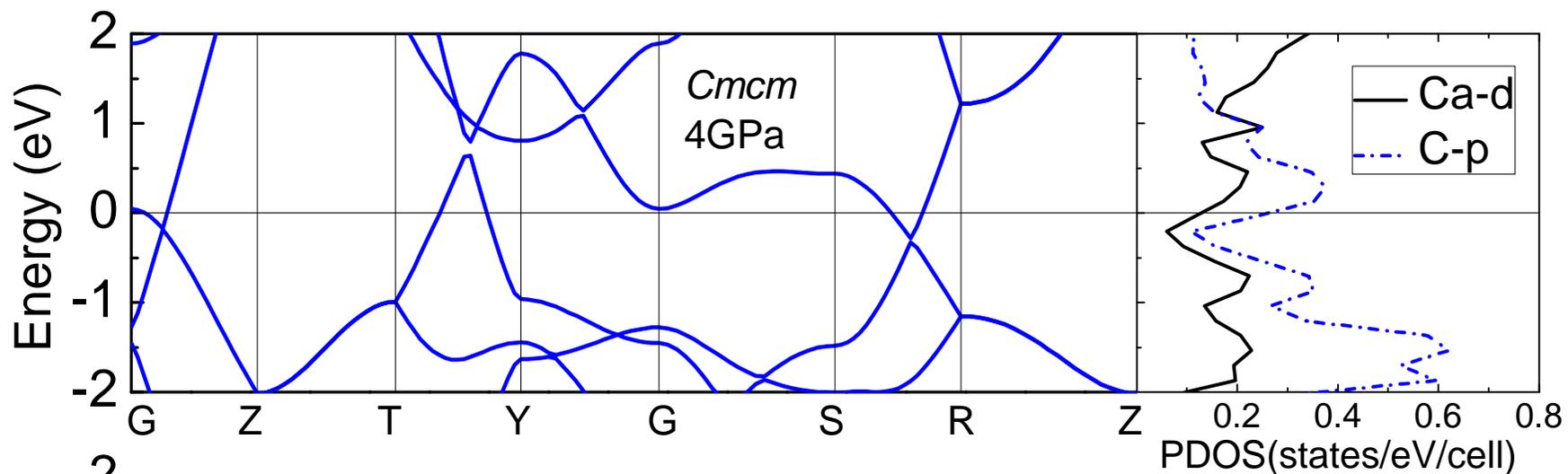
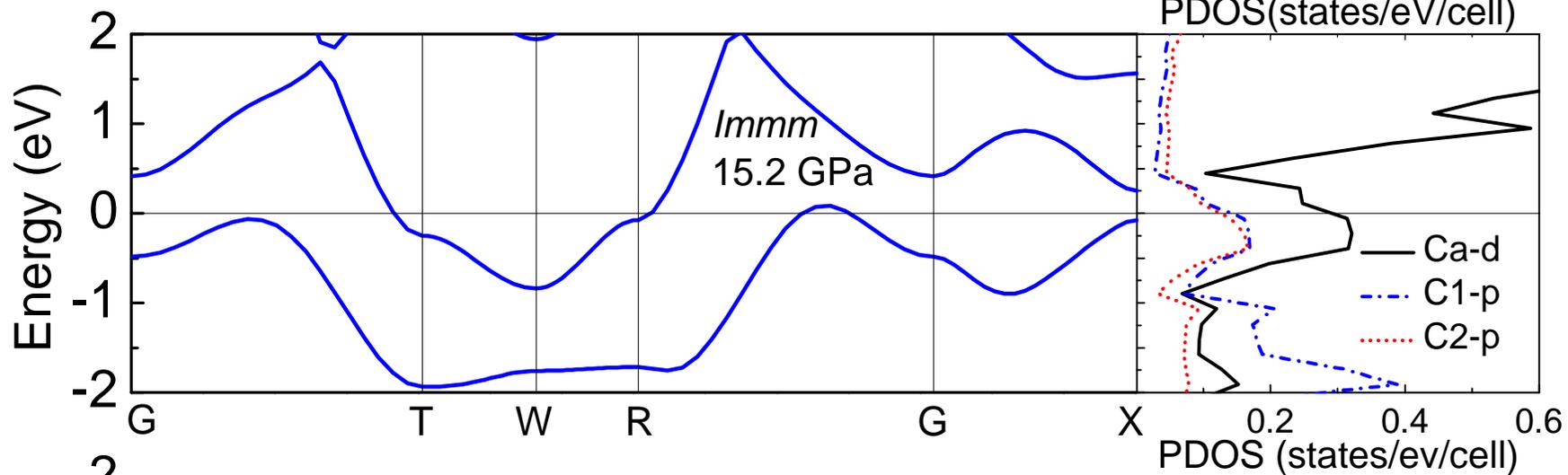
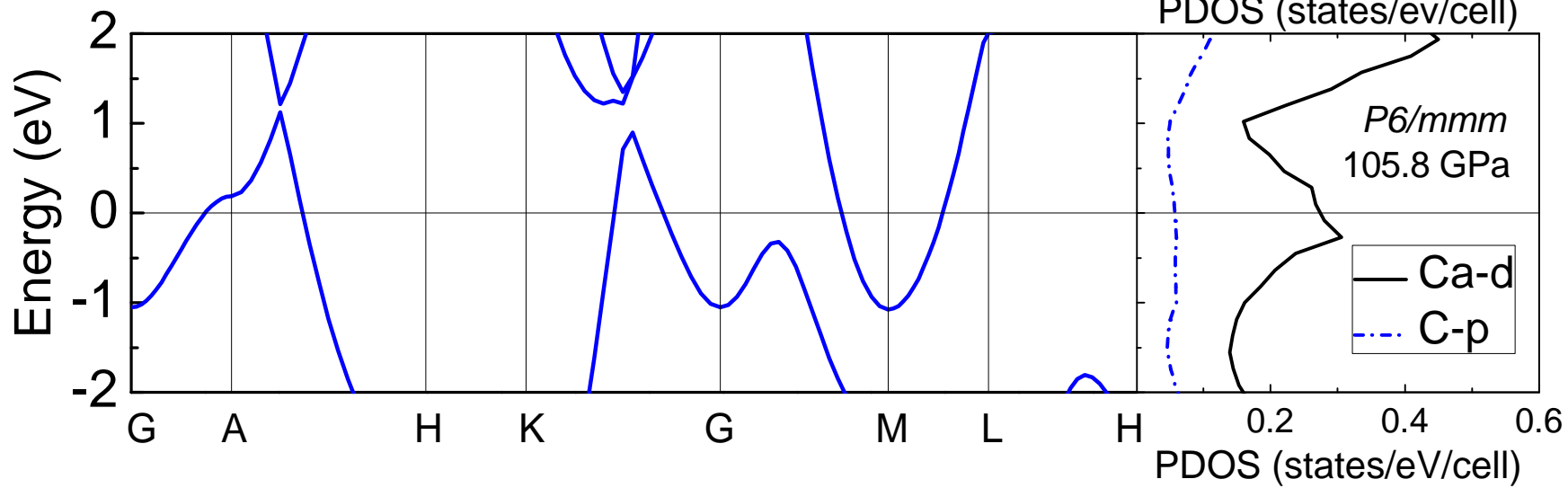

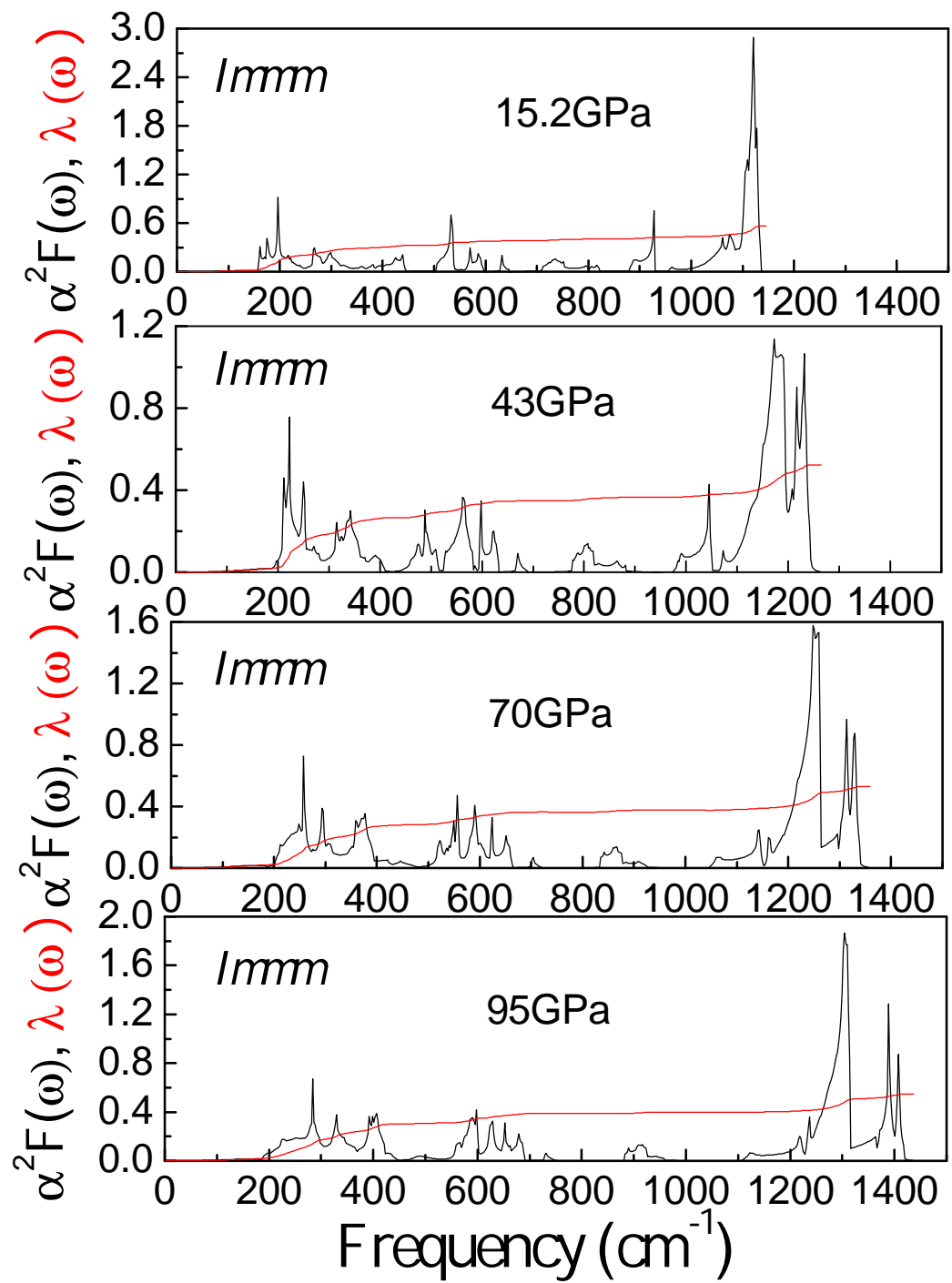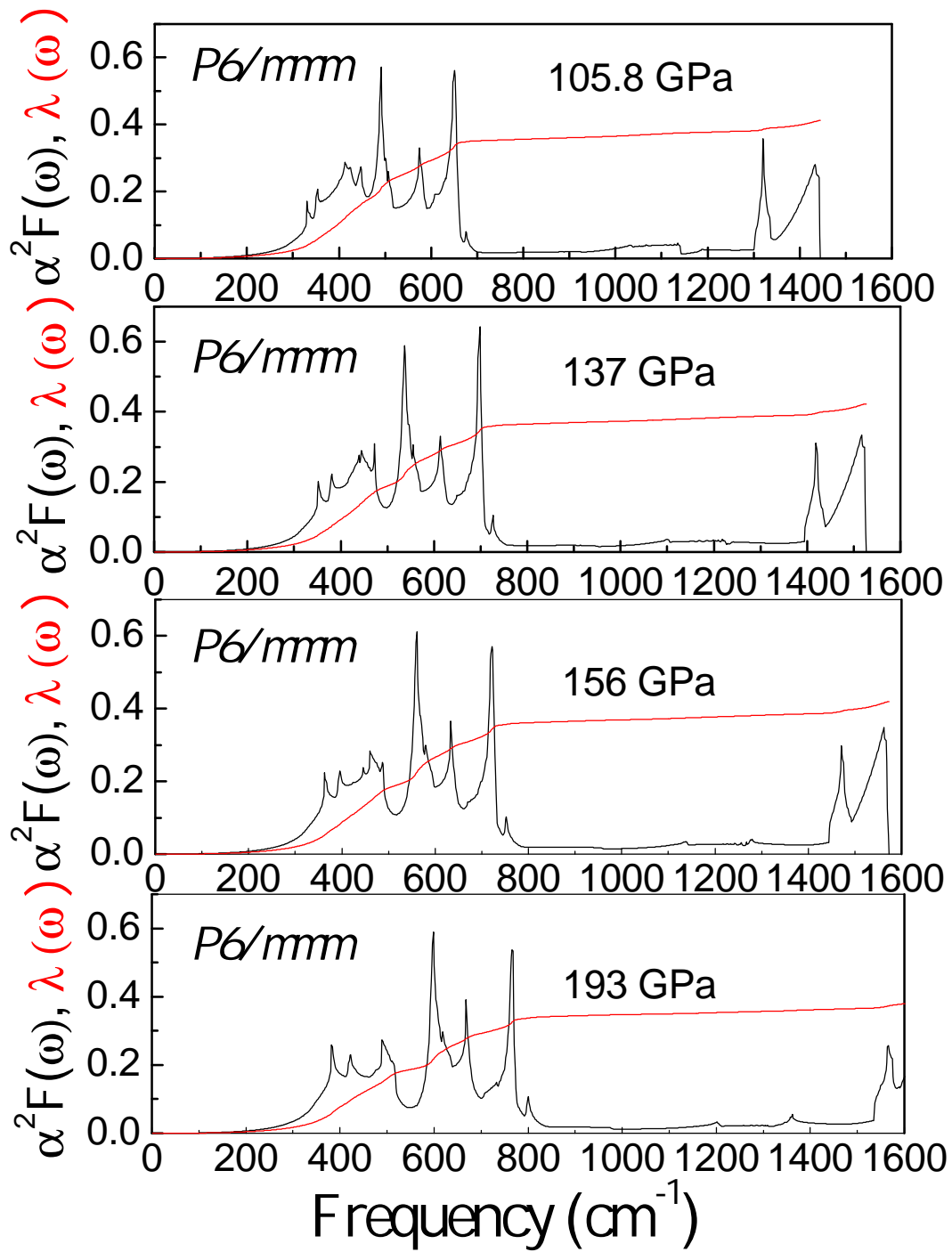

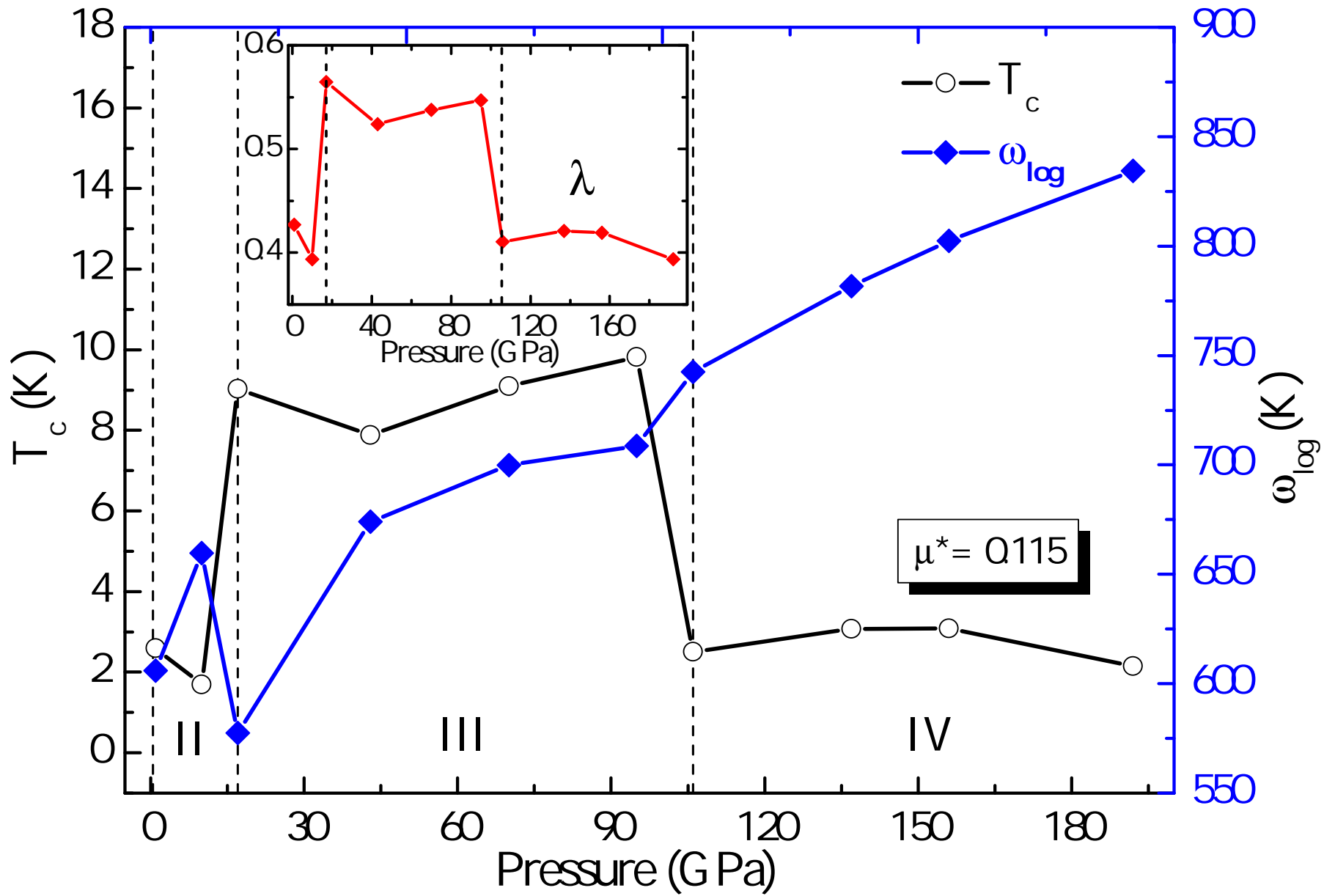

# Pressure-induced Superconductivity in CaC$_2$

Yan-Ling Li,[a,b] Wei Luo,[b,c] Zhi Zeng,[d] Hai-Qing Lin,[e] Ho-kwang Mao,[f] and Rajeev Ahuja[b,c*]

[a] *School of Physics and Electronic Engineering, Jiangsu Normal University, 221116, Xuzhou, P. R. China*

[b] *Condensed Matter Theory Group, Department of Physics and Astronomy, Uppsala University, P.O. Box 516, SE-751 20 Uppsala, Sweden*

[c] *Applied Material Physics, Department of Materials Science and Engineering, Royal Institute of Technology (KTH), SE-100 44, Stockholm, Sweden*

[d] *Key Laboratory of Materials Physics, Institute of Solid State Physics, Chinese Academy of Sciences, Hefei 230031, P. R. China*

[e] *Beijing Computational Science Research Center, Beijing 100089, P. R. China*

[f] *Geophysical Laboratory, Carnegie Institution of Washington, Washington, DC 20015, USA*

## Figure Legends

Figure S1. Energy band of phase *C2/m* at zero pressure.

Figure S2. Band structure of *Immm* phase at different pressure. Red line: 10 GPa; Blue line: 70 GPa.

Figure S3. The Eliashberg phonon spectral function $\alpha^2 F(\omega)$ (blank line) and integrated $\lambda(\omega)$ (red line) for *Cmcm* at 1GPa.

Figure S4. Phonon spectrum of *P6/mmm* structure at 953 GPa from small displacement method via Phonopy code.

Figure S5. Equation of state of *P6/mmm* structure.

Figure S6. Electronic Density of states of P6/mmm phase at different volumes (A$^3$) (200 GPa - 1TPa).

## Tables

Table S1. Structures of the stable phases of CaC$_2$. Only the fractional coordinates of symmetry inequivalent atoms are given.

Table S2. Mulliken Population analysis for phase P6/mmm at 105.8 GPa obtained using CASTEP code.

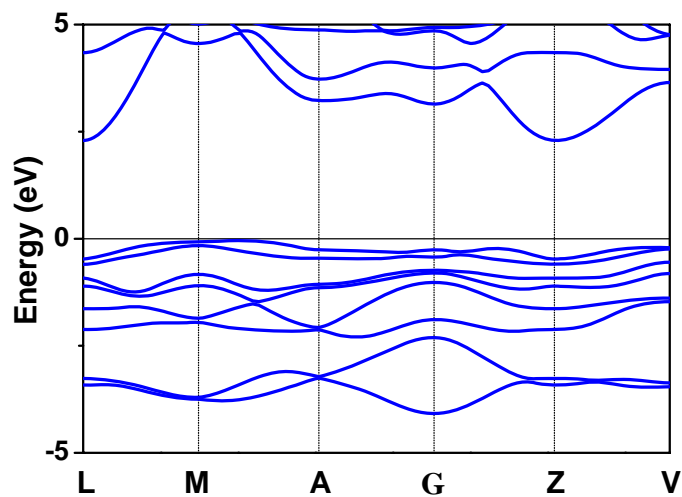

Figure S1. Energy band of phase *C2/m* at zero pressure using PBE exchange-correlation function and PAW pseudopotential as implemented in VASP [1-4].

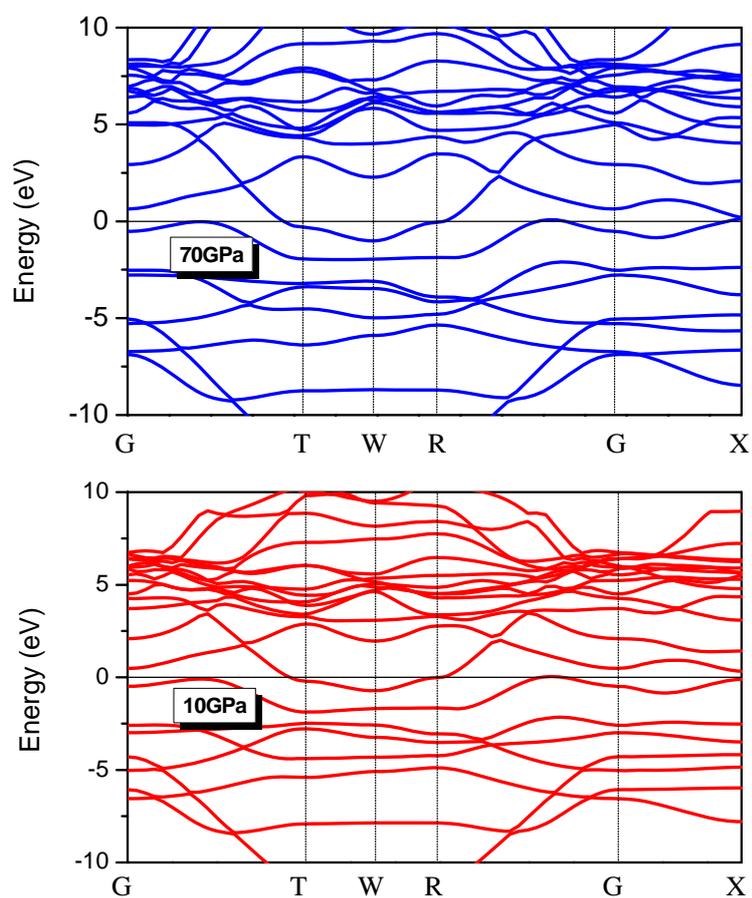

Figure S2. Band structure of *Immm* phase at different pressure. Red line: 10 GPa; Blue line: 70 GPa. Under compression, conduction band shifts to higher energy level, while covalence band shifts to lower level.

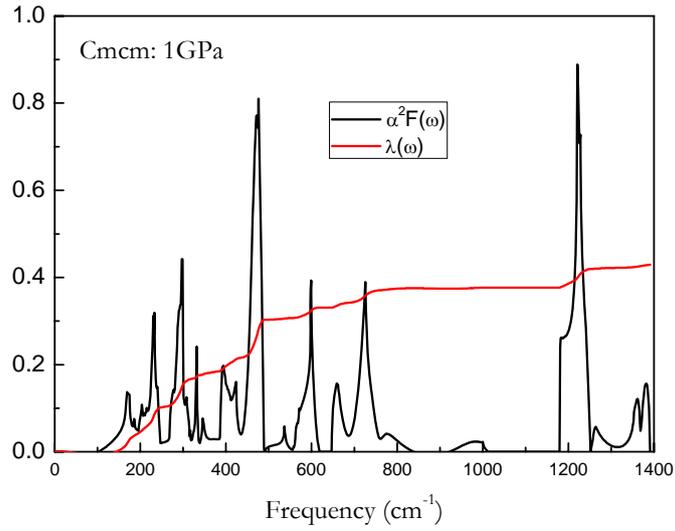

Figure S3. The Eliashberg phonon spectral function $\alpha^2 F(\omega)$ (blank line) and integrated $\lambda(\omega)$ (red line) for *Cmcm* at 1GPa calculated using DFPT as implemented in Quantum Espresso[5].

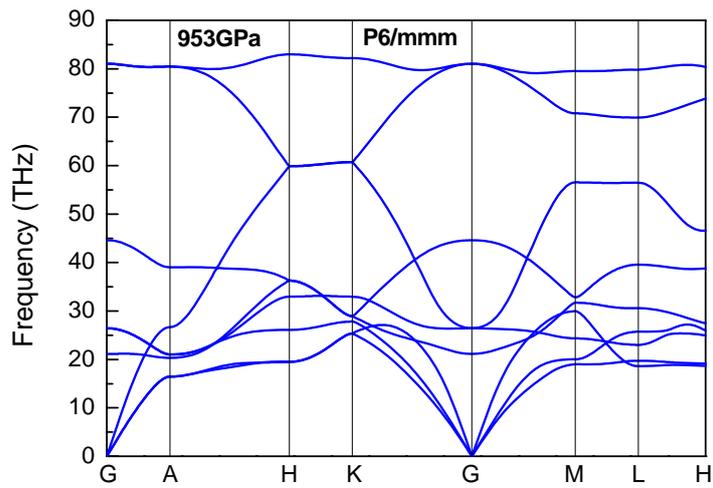

Figure S4. Phonon spectrum of *P6/mmm* structure at 953 GPa from small displacement method via Phonopy code [6].

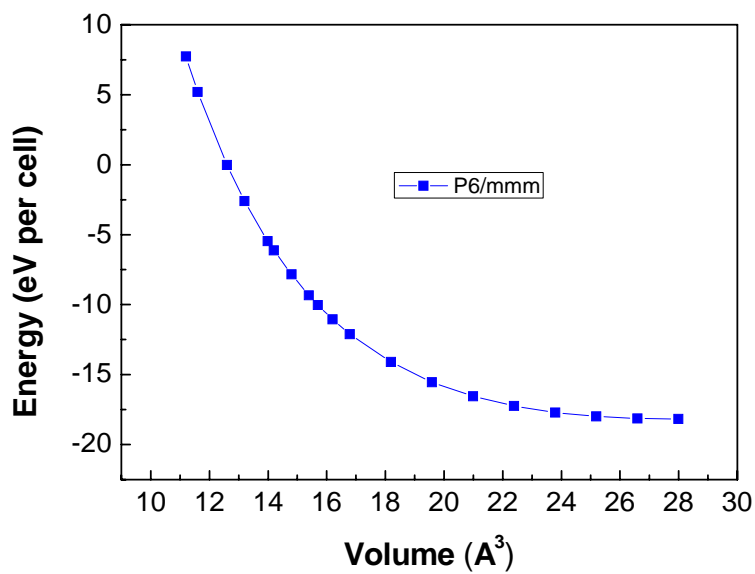

Figure S5. Equation of state of *P6/mmm* structure.

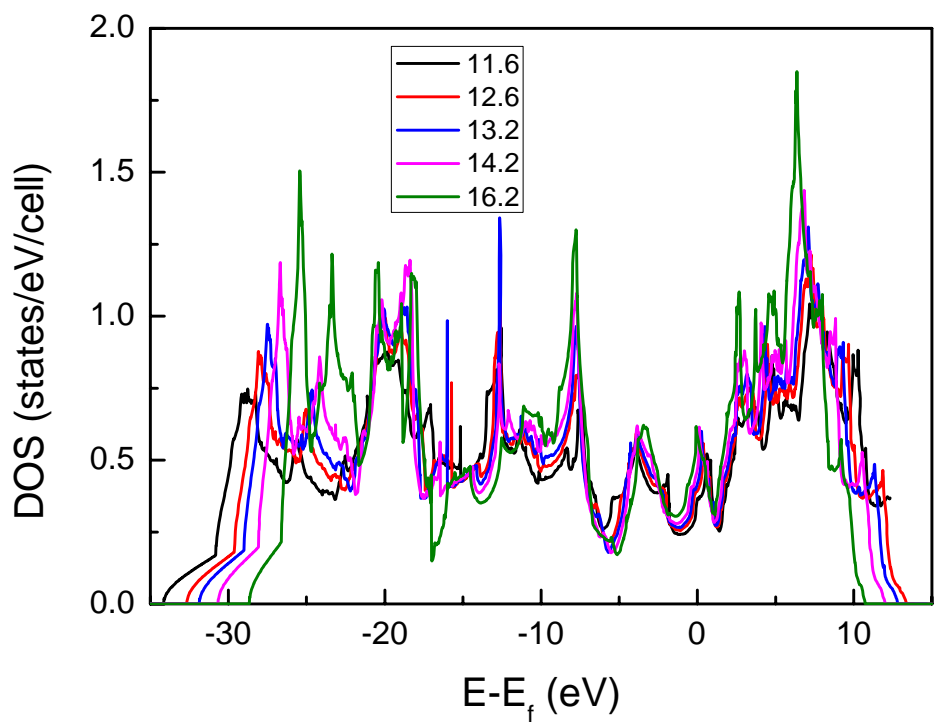

Figure S6. Electronic Density of states (DOS) of *P6/mmm* phase at different volumes ($A^3$) (200 GPa - 1TPa). As pressure increases, DOS value at Fermi level ($E_f$) decreases.

Table SI. Structures of the stable phases of CaC$_2$. Only the fractional coordinates of symmetry inequivalent atoms are given.

| Pressure (GPa) | Space group (No.) | Lattice parameters $(a, b, c, \alpha, \beta, \gamma)$ (Å, °) | | | Atomic fractional coordinates | | | |
|---|---|---|---|---|---|---|---|---|
| 0 | C2/m | 7.1712 | 3.8463 | 8.7162 | Ca 4$i$ | 0.0425 | 0.0000 | 0.2470 |
| | (12) | 90 | 124.99 | 90 | C1 4$i$ | 0.6063 | 0.0000 | 0.0594 |
| | | | | | C2 4$i$ | 0.5142 | 0.0000 | 0.4352 |
| 4 | Cmcm | 3.6822 | 8.6324 | 4.7360 | Ca 4$c$ | 0.0000 | 0.1465 | 0.2500 |
| | (63) | 90 | 90 | 90 | C 8$f$ | 0.000 | 0.4376 | 0.1028 |
| 15.2 | Immm | 7.0623 | 2.6317 | 6.2697 | Ca 4$e$ | 0.2951 | 0.0000 | 0.0000 |
| | (71) | 90 | 90 | 90 | C1 4$i$ | 0.0000 | 0.0000 | 0.2371 |
| | | | | | C2 4$j$ | 0.5000 | 0.0000 | 0.3829 |
| 105.8 | P6/mmm | 2.5412 | 2.5412 | 3.6864 | Ca 1$a$ | 0.0000 | 0.0000 | 0.0000 |
| | (191) | 90 | 90 | 120 | C 2$d$ | 0.3333 | 0.6667 | 0.5000 |

Table S2 Atomic Mulliken population analysis obtained using CASTEP code [7].

| | *s* | *p* | *d* | Total | Charge |
|---|---|---|---|---|---|
| C | 1.11 | 3.29 | - | 4.39 | -0.39 |
| C | 1.11 | 3.29 | - | 4.39 | -0.39 |
| Ca | 1.80 | 5.92 | 1.49 | 9.22 | 0.78 |